\documentclass[aps,twocolumn,superscriptaddress,showpacs]{revtex4-1}

\usepackage{amsmath}
\usepackage{amssymb}
\usepackage{graphicx}
\usepackage{xcolor}
\usepackage{hyperref}
\usepackage{ulem}

\begin{document}

\title{Control of effective free energy landscape in a frustrated magnet by a field pulse}

\author{Yuan Wan}
\affiliation{Perimeter Institute for Theoretical Physics, Waterloo, Ontario N2L 5G7, Canada}
\author{Roderich Moessner}
\affiliation{Max Planck Institute for the Physics of Complex Systems, 01187 Dresden, Germany}

\begin{abstract}

Thermal fluctuations can lift the degeneracy of a ground state manifold, producing a {\it free energy} landscape without accidentally degenerate minima. In a process known as order by disorder, a subset of states incorporating symmetry-breaking may be selected. Here, we show that such a free energy landscape can be controlled in a non-equilibrium setting as the slow motion within the ground state manifold is governed by the fast modes out of it. For the paradigmatic case of the classical pyrochlore XY antiferromagnet, we show that a uniform magnetic field pulse can excite these fast modes to generate a tunable effective free energy landscape with minima at thermodynamically unstable portions of the ground state manifold. 
\end{abstract}

\date{\today}


\maketitle

The emergence of a thermodynamic landscape from microscopic interactions, and its consequences for macroscopic behavior, is a central theme of condensed matter physics. The essential role of fluctuation contributions to the free energy landscape is underlined by entropic interactions in soft matter~\cite{Kirkwood1939,Guth1941,Onsager1949}, infinite temperature phase transitions~\cite{Alder1957,Stillinger1967}, and, at low temperatures, the phenomenon of order by disorder (ObD)~\cite{Villain1980,Shender1982,Kawamura1984,Henley1989,Moessner1998}.

This raises the question of how one can control a free energy landscape in a condensed matter system by manipulating fluctuations instead of changing its Hamiltonian. In equilibrium (EQ), the Boltzmann distribution fully dictates the free energy landscape and hence allows little room for control. In this work, we show that such control is feasible if one drives the system out of equilibrium, where fluctuations dynamically produce a free energy landscape which may be tuned by changing the non-equilibrium conditions. 

To illustrate this, we sketch a simple but generic scenario. It is based on a crisp distinction between energetic and entropic contributions to the free energy as exists at low energies in a broad class of geometrically frustrated magnets. This arises because geometrical frustration often results in an accidentally degenerate ground state (GS) manifold, i.e. a large, continuous family of degenerate GSs not related by any symmetry operation. In EQ, the entropy due to the thermal fluctuations near each GS can vary along the manifold. Parametrizing the ground state manifold by coordinates $\mathcal{Q}$, one thus obtains an effective free energy $V(\mathcal{Q})$ from integrating out the fluctuations, which can thus lift the accidental degeneracy and tend to stabilize the GS(s) with maximal entropy.  

\begin{figure}
\includegraphics[width=0.4\columnwidth]{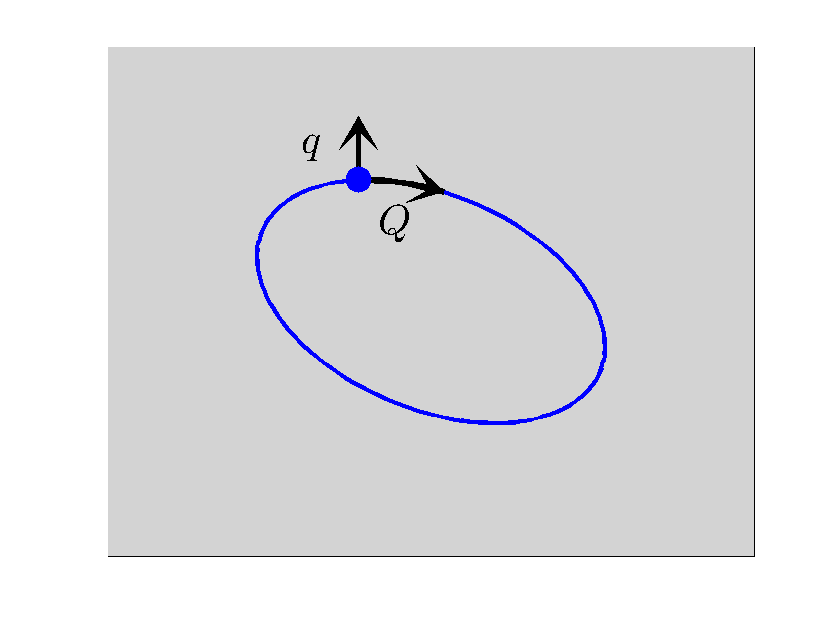}
\caption{(Color online) Geometrical frustration for vector spins produces an accidentally degenerate GS manifold (blue ellipse) in many-body configuration space (gray box). $\mathcal{Q}$ parametrizes the GS manifold, and $q$ the fluctuations out of it.}
\label{fig:sketch}
\end{figure}

We focus on systems with Hamiltonian dynamics such as frustrated vector spin models. We separate variables into the pseudo-Goldstone modes describing the drift motion within the GS manifold, and the other normal modes involving deviations out of it (Fig.~\ref{fig:sketch})~\cite{Moessner2001}. In the low temperature limit, the drift motion within GS manifold is vanishingly slow as they experience no linear restoring force. By contrast, the other modes  are fast owing to their finite stiffness.  Integrating these out (a controlled procedure provided their frequencies are bounded above zero) yields a non-equilibrium contribution to the effective free energy of the form $V(\mathcal{Q})\propto\sum_{i}I_{i}\sqrt{K_i(\mathcal{Q})}$, with action variable $I_i>0$ of the $i$-th excited normal mode, and $K_i$ its $\mathcal{Q}$-dependent stiffness~\cite{Doucot1998}. Importantly, $V(\mathcal{Q})$ can be tuned by adjusting the weights $I_i$. This we show can be achieved straightforwardly by a magnetic field pulse.

\begin{figure}
\includegraphics[width=\columnwidth]{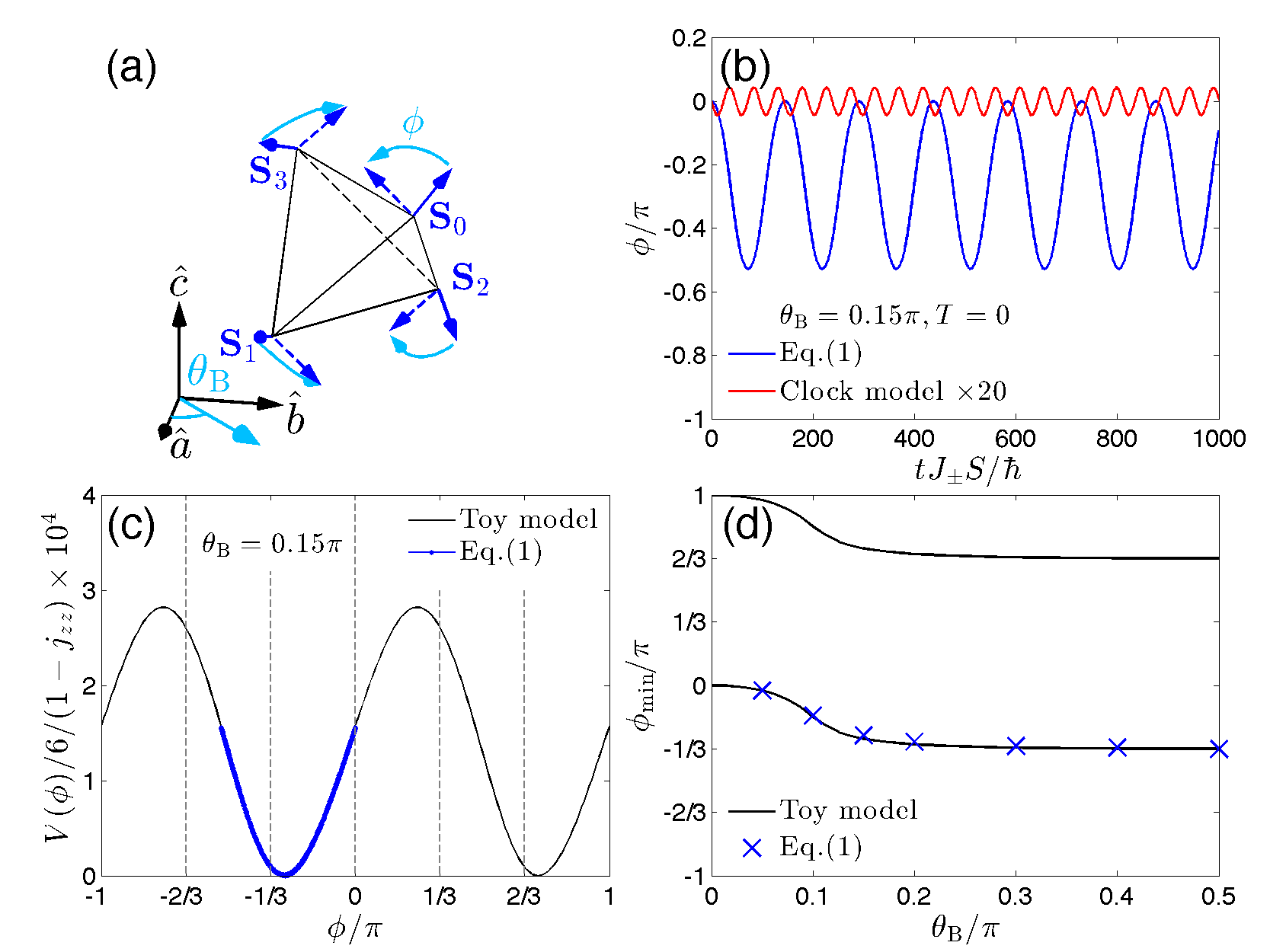}
\caption{(Color online) (a) GS of Eq.~\eqref{eq:model} in a unit cell. $0\sim3$ label the sublattices. Starting from a GS (blue solid arrows, corresponding to $\phi=0$), one obtains another GS (blue broken arrows, $\phi=\pi/2$) by rotating all spins with respect to their local 3-fold axes. Inset: Crystallographic axes (black arrows), pulse polarization (light blue arrow), and polarization angle $\theta_\mathrm{B}$. (b) GS N\'{e}el order parameter argument $\phi$ as function of time $t$ after a pulse with polarization angle $\theta_\mathrm{B}=0.15\pi$. Results from the simulation of Eq.~\eqref{eq:model} (blue) contrasted to a model with an inherently energetic (rather than emergent entropic) clock term (red). (c) Effective potential energy density $V(\phi)/(6(1-j_{zz}))$ in the GS manifold from the toy model (black) and the model simulation in (b) (blue). Gray dashed lines mark the positions of $\psi_2$ states. (d) Potential minima $\phi_\mathrm{min}$ as function of pulse polarization angle $\theta_\mathrm{B}$. Black solid lines show the two degenerate minima predicted by the toy model. Blue crosses mark the center position of the $\phi$ oscillation extracted from simulation.}
\label{fig:tzero}
\end{figure}

We flesh out this scenario with a concrete model, for which we establish properties of the effective landscape, its tunability and resulting dynamics, as well as its interplay with the thermal fluctuations. We also develop an analytic toy model that transparently explains the central phenomenon. 

Consider the classical pyrochlore XY antiferromagnet, which has gained prominence thanks to its likely realization of ObD in the rare earth magnet Er$_2$Ti$_2$O$_7$~\cite{Champion2003,Champion2004,Zhitomirsky2012,Savary2012,Wong2013,Oitmaa2013,McClarty2014,Zhitomirsky2014}:
\begin{align}
H = &\sum_{\langle ij\rangle}-\frac{J_\pm}{2}(S^+_iS^-_j+h.c.)+\frac{J_{\pm\pm}}{2}(e^{i\psi_{ij}}S^+_iS^+_j+h.c.)\nonumber\\
& -J^{zz}S^z_iS^z_j.
\label{eq:model}
\end{align}
Here, spin $\mathbf{S}_i$ of length $S$ resides on pyrochlore lattice site $i$. The summation runs over nearest-neighbor bonds. $S^{x,y,z}_i$ are the Cartesian components of the spin in the local frame $\{\hat{x}_i,\hat{y}_i,\hat{z}_i\}$, and $S^\pm_i \equiv S^x\pm iS^y_i$. $\psi_{ij}$ are link-dependent phase angles due to local spin frames (See \cite{SM} for details). $J_{\pm},J_{zz}$ parametrize anisotropic Heisenberg exchange interaction, whereas $J_{\pm\pm}$ originates from a Dzyaloshinskii-Moriya or pseudo-dipolar interaction. We assume $J_{\pm}>J_{\pm\pm},J_{zz}>0$. For simplicity, we have omitted a symmetry-allowed term known as $J_{z\pm}$ as it will not change the physics discussed in this work.

\begin{figure}
\includegraphics[width=\columnwidth]{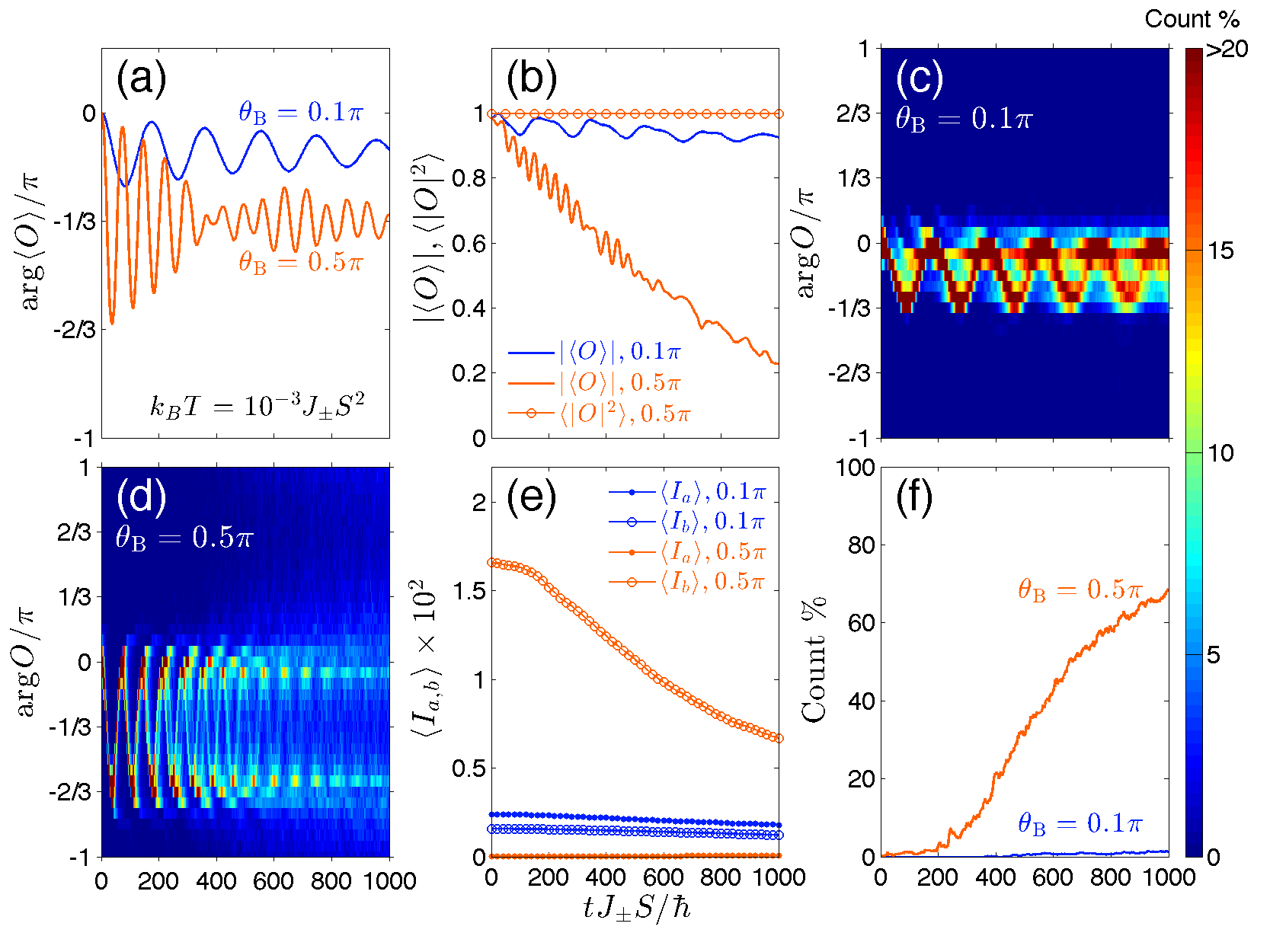}
\caption{(Color online)  (a) Argument and (b) modulus of the ensemble-averaged complex N\'{e}el order parameter $\langle O\rangle$ as a function of time $t$ at $k_BT=10^{-3}J_{\pm}S^2$. Blue and orange lines are for pulse polarization angles $\theta_\mathrm{B}=0.1\pi$ and $0.5\pi$, respectively. (b) also shows the ensemble average of modulus squared, $\langle |O|^2\rangle$, for $\theta_\mathrm{B}=0.5\pi$ (orange open circles). (c)(d) Histogram of the N\'{e}el order parameter argument $\mathrm{arg}\,O$ as function of time $t$ at the same temperature. (e) Same as (a) but for ensemble average of action variables $I_{a}$ (dots) and $I_{b}$ (open circles) associated with the optical magnons $a$ and $b$. (f) Number of trajectories for which the total energy of the pseudo-Goldstone mode exceeds the potential maxima. }
\label{fig:tfinite}
\end{figure}

We briefly review the ObD following from Eq.~\eqref{eq:model} \cite{Zhitomirsky2012,Savary2012,Wong2013,Oitmaa2013,McClarty2014,Zhitomirsky2014}. The GSs, which preserve lattice translation symmetry, show N\'{e}el order with complex order parameter $O\equiv\sum_i S^+_i/(\mathcal{N}S)$ ($\mathcal{N}$ is the number of sites). GS spin configurations are $\mathbf{S}_i = S(\cos\phi\hat{x}_i+\sin\phi\hat{y}_i)$ with $\phi=\arg O$ (Fig.~\ref{fig:tzero}a). Crucially, the GS energy is independent of $\phi$: Eq.~\eqref{eq:model} possesses an accidental $U(1)$ degeneracy although it only exhibits discrete symmetries.

At small but finite temperature, the entropy due to the spin wave fluctuations lifts the accidental degeneracy, yielding six symmetry-equivalent maxima known as $\psi_2$ states, located at $\phi=m\pi/3$, $m\in\mathbb{Z}$: thermal ObD corresponds to an emergent entropic six-state clock anisotropy in the $U(1)$ manifold.

To study the non-EQ dynamics of Eq.~\eqref{eq:model}, we endow the spins with precessional Landau-Lifshitz (LL) dynamics. We begin with a toy model that captures the essential features: taking the spins on each of the four sublattices $i=0\sim3$ with the same orientation, $\mathbf{S}_i/S = \sqrt{1-z^2_i}(\cos\phi_i\hat{x}+\sin\phi_i\hat{y})+z_i\hat{z}_i$ reduces the degrees of freedom to the four spins in a unit cell. In other words, we freeze spin wave modes with wave vector $\mathbf{k}\neq0$.

We define $\phi \equiv (\phi_0+\phi_1+\phi_2+\phi_3)/4$, $\phi_a \equiv (\phi_0+\phi_1-\phi_2-\phi_3)/2$,  $\phi_b \equiv (\phi_0+\phi_2-\phi_1-\phi_3)/2$, and $\phi_c \equiv (\phi_0+\phi_3-\phi_1-\phi_2)/2$, along with corresponding $z$ and $z_{a,b,c}$. $\phi$ corresponds to the pseudo-Goldstone mode, whereas $\phi_{a,b,c}$ correspond to finite-frequency optical magnons. Linearizing Eq.~\eqref{eq:model} in $\phi_{a,b,c}$ yields \cite{SM}:
\begin{align}
\ddot{\phi}_{l} = -\omega^2_0 f_{l}(\phi)\phi^2_{l};\,\,
\ddot{\phi} =-\widetilde{\omega}^2_0 \sum_{l=a,b,c}\phi^2_l\sin(2\phi+\Theta_l),
\label{eq:eom_toy}
\end{align}
with rescaled variables such that $J_\pm S\to 1$, $j_{\pm\pm}\equiv J_{\pm\pm}/J_{\pm}$, $j_{zz} \equiv J_{zz}/J_{\pm}$. $l$ runs over labels $a,b,c$. $f_{l}(\phi) = 1-j_{\pm\pm}/2\,\cos(2\phi+\Theta_{l})$. $\Theta_{a} \equiv 0$, $\Theta_b \equiv 2\pi/3$, and $\Theta_c \equiv -2\pi/3$. $\omega_0\equiv4\sqrt{3+j_{zz}}$ and $\widetilde{\omega}_0\equiv\sqrt{6j_{\pm\pm}(1-j_{zz})}$ are constant frequencies. 

Since $\phi_{l}\ll1$, the optical magnons are fast harmonic oscillators parametrically driven by the slow motion of the pseudo-Goldstone mode $\phi$. We thus proceed to integrate out optical magnons by using the method of averaging~\cite{Arnold1989,Doucot1998}. We obtain~\cite{SM}: $\ddot{\phi} = -\partial{}V(\phi)/\partial\phi$, where
\begin{align}
V(\phi) = \frac{3(1-j_{zz})}{\sqrt{3+j_{zz}}}\sum_{l=a,b,c}I_l\sqrt{f_l(\phi)},
\label{eq:veff}
\end{align}
is the effective potential in the $U(1)$ degenerate manifold. $I_{l}$, the action variable of mode $l$, is an adiabatic invariant~\cite{Landau1976}, so that $V(\phi)$ is approximately time independent.

For $I_a=I_b=I_c$, $V(\phi) \propto -j^3_{\pm\pm} \cos(6\phi)+o(j^3_{\pm\pm})$ with minima at the states selected by ObD in EQ. For generic values of $I_{a,b,c}$, however, the
 minima are two-fold degenerate, $V(\phi)=V(\phi+\pi)$ and, crucially,  located elsewhere. 

Selectively exciting optical magnons thus permits control of the individual values of $I_{a,b,c}$ and thereby $V(\phi)$. This can in fact be achieved simply via the polarization of an applied magnetic field pulse. To see this, note that the magnetization $M_{a,b,c}$ along the crystallographic $a,b,c$ axes, to leading order in $\phi_{a,b,c}$ and $z_{a,b,c}$ is \cite{Maryasin2016,SM},
\begin{align}
\frac{M_{a,b,c}}{\mathcal{N}\mu_\mathrm{B}S} = -\frac{2g_\parallel}{\sqrt{3}}z_{a,b,c}-\frac{4g_\perp}{\sqrt{6}}\sin(\phi-\Theta_{a,b,c})\,\phi_{a,b,c}\ ,
\label{eq:magnetization}
\end{align}
with $g_\parallel (g_\perp)$ being the Land\'{e} g-factor along the local $\hat{z}$ axis ($\hat{x},\hat{y}$ axes): the $a,b,c$ magnons respectively carry magnetic dipole moments in the three crystallographic axes. 

We next confirm numerically such generation and control of  an effective free energy landscape. We first equilibrate the model at temperature $T$ and then apply a short magnetic pulse whose temporal profile is a Dirac-$\delta$ function: $\mathbf{B}(t)=B_\mathrm{max}\hat{n}\delta(t/\tau)$, where $B_\mathrm{max}$ is the peak strength, $\tau$ is the duration, and $\hat{n}$ is the polarization. After the pulse, we remove the bath and let the system evolve according to the LL equation at $t>0$. 

The simulation uses a periodic lattice of $8\times8\times8$ unit cells. We do not observe significant system size dependence in dynamics~\cite{SM}. We integrate the LL equation via the 4th order Runge-Kutta method (RK4). For nonzero $T$, we generate $2^{10}$ initial states for the LL equation from canonical Monte Carlo (MC) and average over microcanonical trajectories. The RK4 step width is chosen such that the relative error of energy $\epsilon<10^{-5}$. The integration stops at $10^3\hbar/(J_{\pm}S)$, corresponding to $>10^3$ oscillation cycles of optical magnons. Model parameters $j_{\pm\pm} = 0.646$, $j_{zz} = 0.192$, $g_\parallel = 2.45$, and $g_\perp = 6.0$ are similar to those for Er$_2$Ti$_2$O$_7$ \cite{Savary2012}. In EQ, the $\psi_2$ states occur with equal probability. We assume the model is initially in a single domain with $\phi=0$. Only the area of the $\delta$ peak $B_\mathrm{max}\tau$ enters the equation of motion, which we set to 0.1 ps$\cdot$T~\cite{SM}. The pulse polarization lies in the $ab$ plane, i.e. $\hat{n}=\cos\theta_\mathrm{B}\hat{a}+\sin\theta_\mathrm{B}\hat{b}$ (Fig.~\ref{fig:tzero}a, inset). In this setup, the pulse only excites $a$ and $b$ magnons. The energy deposited by the pulse is $\sim10^{-3}\,J_{\pm}S^2$ per spin.

First consider the $T=0$ limit. $\phi$ initially rests at $0$ for $t<0$. The pulse generates a $V(\phi)$ whose minima are located elsewhere. $\phi$ thus oscillates around a nearby minimum of $V(\mathrm{\phi})$ (see Fig.~\ref{fig:tzero}b for $\theta_\mathrm{B}=0.15\pi$). To extract $V(\phi)$ from simulation, we use the conservation of energy~\cite{SM}, $V(\phi)/(6(1-j_{zz})) + K = \mathrm{const.}$, where the first and second terms are respectively the potential and kinetic energy density of the pseudo-Goldstone mode. $K$ can be evaluated from data. The result is in good agreement with Eq.~\eqref{eq:veff}, with $I_{a,b,c}$ extracted from the initial condition (Fig.~\ref{fig:tzero}c).

\begin{figure}
\includegraphics[width=\columnwidth]{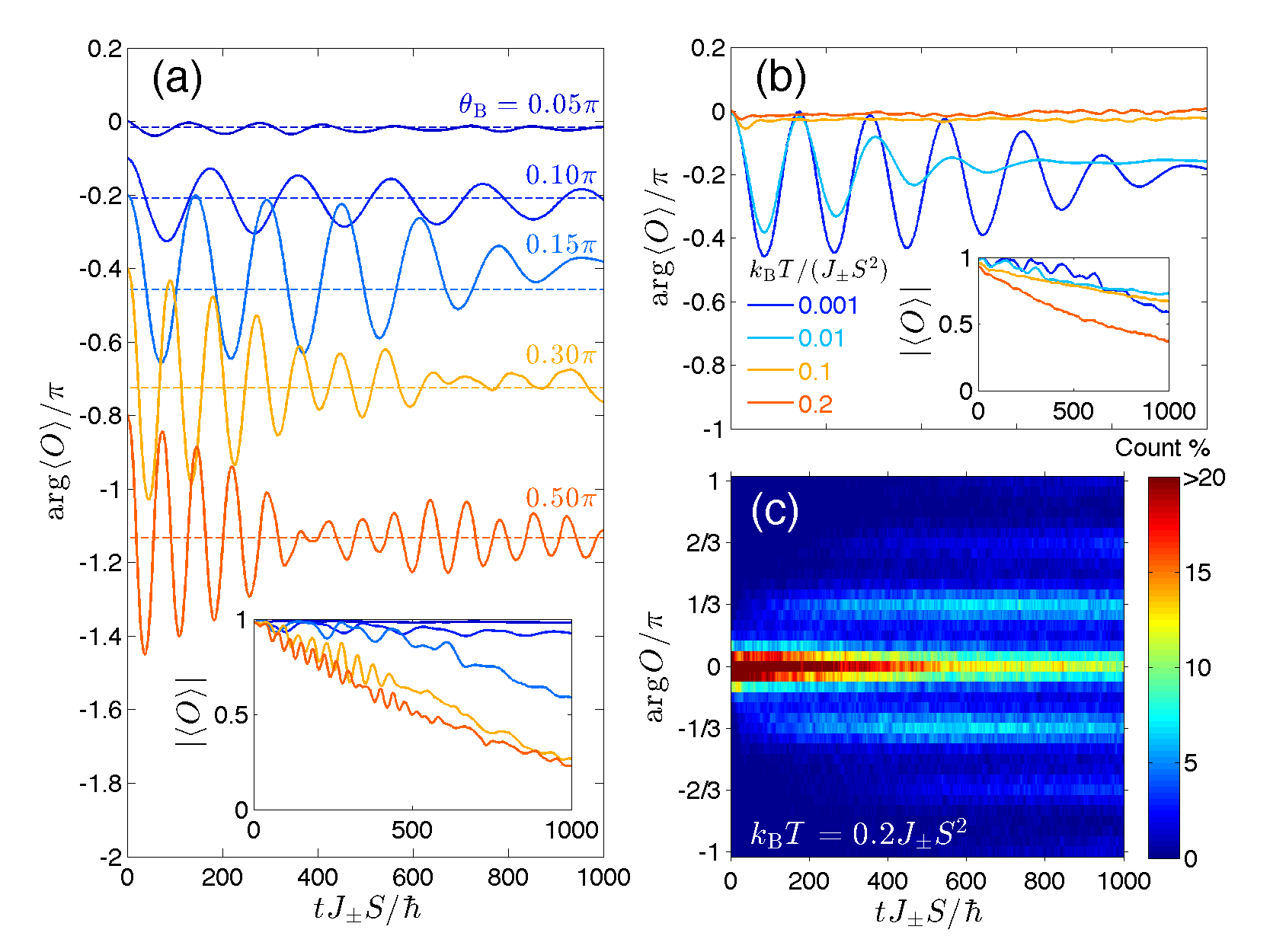}
\caption{(Color online) (a) Argument and modulus (inset) of the ensemble-averaged  N\'{e}el order parameter $\langle{}O\rangle$ versus time $t$ for various pulse polarizations at $k_BT=10^{-3}J_{\pm}S^2$. $\mathrm{arg}\langle{}O\rangle=0$ at time $t=0$, but plots are shifted vertically for visibility. Dashed lines mark the predicted effective potential minima from the toy model. (b) Same as (a) but for fixed pulse polarization angle $\theta_\mathrm{B}=0.15\pi$ at four different temperatures. (c) Histogram of N\'{e}el order parameter argument $\mathrm{arg}\,O$ as function of time $t$ for $\theta_\mathrm{B}=0.15\pi$ at $k_BT=0.2J_{\pm}S^2$.}
\label{fig:scan}
\end{figure}

We next show that the accidental degeneracy underpinning ObD is constitutive to the controllability of the free energy landscape in our present scheme. To do this, we contrast Eq.~\eqref{eq:model} with a clock model where ObD is mimicked by microscopic interactions yielding the same degeneracy lifting~\cite{Rau2016}: we add a six-state clock anisotropy term, $-\Delta/S^4[(S^+_i)^6+h.c.]$, to Eq.~\eqref{eq:model} and set $j_{\pm\pm}=0$. For the clock model, $V(\phi) \propto -\Delta\cos(6\phi)$, and the pulse {\it only produces a small renormalization of $\Delta$}:  for $\Delta/J_{\pm} = 10^{-4}$, $\phi$ oscillates around $0$ (Fig.~\ref{fig:tzero}b). The oscillation amplitude is small as the N\'{e}el order couples to the magnetic field only nonlinearly. 
 
We further demonstrate the control on $V(\phi)$ by scanning the field polarization $\hat{n}$. As $\hat{n}$ rotates from the $a$ to the $b$ axis, $I_b$ increases while $I_a$ decreases. The $I_a$ term of Eq.~\eqref{eq:veff} favors $\phi = 0,\pi$, whereas the $I_b$ term favors $\phi=-\pi/3,2\pi/3$. Thus, the minimum positions $\phi_\mathrm{min}$ continuously shift from $0,\pi$ to $-\pi/3, 2\pi/3$. In numerical simulation, the shift of $\phi_\mathrm{min}$ is manifest as the change of the center position of $\phi$ oscillation (Fig.~\ref{fig:tzero}d). The small discrepancy between the toy model and the simulation is likely due to nonlinear effects neglected in Eq.~\eqref{eq:eom_toy}.

At nonzero temperature, the effective potential produced by optical magnons is also subject to the thermal fluctuations in $\mathbf{k}\neq0$ modes. In EQ, these modes produce an entropic effective potential with the magnitude of $\sim10^{-3}\,k_BT$ per spin. We now show that the above non-EQ phenomena are nonetheless robust at low $T$. We start with $k_BT/(J_{\pm}S^2)=10^{-3}$. Fig.~\ref{fig:tfinite}a,b show two representative cases. For $\theta_\mathrm{B}=0.1\pi$, $\mathrm{arg}\langle{}O\rangle$, the argument of the average N\'{e}el order parameter, exhibits a persistent oscillation with small damping. On the other hand, for $\theta_\mathrm{B}=0.5\pi$, in addition to oscillation in the argument, the modulus $|\langle{}O\rangle|$ decreases quickly. However, the N\'{e}el order persists as the average modulus squared $\langle |O|^2\rangle\approx1$ within the simulation time window (Fig.~\ref{fig:tfinite}b). Thus, the decrease in $|\langle O\rangle|$ is due to the loss of anisotropy, i.e.\ the argument of $O$ explores the $U(1)$ degenerate manifold. This is made clear by the histogram of $\mathrm{arg}\,O$ for $\theta_\mathrm{B}=0.5\pi$, which spreads out in the $U(1)$ manifold at late time (Fig.~\ref{fig:tfinite}d).

The difference between these two types of behavior originates dynamically as follows. Consider the $a$ and $b$ magnons underpinning $V(\phi)$. Their ensemble-averaged action variables $\langle I_{a,b}\rangle$ gradually decay in time, indicating that the energy is being slowly transferred to other modes (Fig.~\ref{fig:tfinite}e). Note $I_{a,b,c}$ must equalize and $\phi$ must return to $\psi_2$ positions after the thermal equilibration time $\tau_\mathrm{eq}$. Yet, the persistent difference in $I_{a,b}$ indicates $\tau_\mathrm{eq}>10^3\hbar/{J_{\pm}S}$. Comparing $\langle{}I_{a,b}\rangle$ for $\theta_\mathrm{B}=0.1\pi$ with $\theta_\mathrm{B}=0.5\pi$, the latter exhibits more pronounced decrease. This slow decay implies that of $V(\phi)$ as well (Eq.~\eqref{eq:veff}). Now, the motion of the pseudo-Goldstone mode $\phi$ is oscillatory as long as its total energy $E$ is less than the potential maxima $V_\mathrm{max}$. As $V_\mathrm{max}$ decreases, $\phi$ may therefore overcome the potential barrier and enter an open orbit. As a qualitative test of this picture, we extract $E$ and $V_\mathrm{max}$ from data and count the number of trajectories for which $E>V_\mathrm{max}$ (Fig.~\ref{fig:tfinite}f). For $\theta_\mathrm{B}=0.1\pi$, the number of such trajectories is negligible, whereas the count steadily grows in time for $\theta_\mathrm{B}=0.5\pi$. 

Having gained a qualitative understanding the two representative cases, we consider the systematic $\theta_\mathrm{B}$-dependence of the order parameter dynamics (Fig.~\ref{fig:scan}a). Throughout, $\mathrm{arg}\langle{}O\rangle$ exhibits damped oscillation, while  $|\langle{}O\rangle|$ decreases at an larger rate as $\theta_\mathrm{B}$ approaches $\pi/2$. Similar to $T=0$, the center of $\mathrm{arg}\langle{}O\rangle$ oscillation gradually shifts from 0 to $-\pi/3$. The center position at early times agrees with the potential minima of Eq.~\eqref{eq:veff} with $I_{a,b,c}$ extracted from initial conditions. The small drift in oscillation center at late time is due to a small change in the relative weight of $I_{a,b,c}$ and the increasing importance of thermal fluctuations. The higher oscillation frequency at larger $\theta_\mathrm{B}$ results from the larger curvature of $V(\phi)$. The beating of the $\mathrm{arg}\langle{}O\rangle$ observed at $\theta_\mathrm{B}=0.5\pi$ is likely due to a finite-width distribution of the oscillation frequencies.

Finally, we study the temperature dependence of this phenomenon. Fig.~\ref{fig:scan}b shows the order parameter dynamics for fixed $\theta_\mathrm{B}=0.15\pi$. Damping of the oscillation in $\mathrm{arg}\langle{}O\rangle$ increases with temperature $T$, while the oscillation center moves toward 0. Both  signify the growing influence of thermal fluctuations. At $k_BT/(J_{\pm}S^2) = 0.2$, $\mathrm{arg}\langle{}N\rangle$ is nearly stationary and very close to $0$, which suggests the thermal fluctuations dominate over the non-EQ potential from optical magnons. This is clearly seen in the histogram of $\mathrm{arg}\,O$ (Fig.~\ref{fig:scan}c): initially at a single $\psi_2$ state, the system escapes to other $\psi_2$ states at late time. Since the model remains XY-ordered at this temperature and hence no symmetry changes occur, we infer a crossover at temperature $T^\ast$, above which the landscape essentially reverts to being thermal. Simple dimensional analysis indicates $T^\ast$ scales with the energy density deposited by the pulse. 

Looking ahead, our identification of a tunable effective free energy landscape leads to a number of interesting directions for future research. Firstly, a more detailed analysis of the finite temperature dynamics is needed for a complete picture of the aforementioned crossover. Secondly, while we have focused on classical models, it is intriguing to see to what extent the physics discussed here has a natural analogue in the quantum realm. This would be directly applicable to quantum magnets such as Er$_2$Ti$_2$O$_7$ and NaCaCo$_2$F$_7$~\cite{Krizan2015,Ross2016,Sarkar2016}, which hold the promise of allowing a detailed and quantitative study of thermal and/or quantum ObD effects.

\begin{acknowledgments}
Research at Perimeter Institute is supported by the Government of Canada through the Department of Innovation, Science and Economic Development Canada and by the Province of Ontario through the Ministry of Research, Innovation and Science. Research at MPI-PKS is  in part supported by DFG under grant SFB 1143. 
\end{acknowledgments}

\bibliography{obd}

\onecolumngrid

\newpage

\appendix

\section{Local spin frames and phase factors}

We define the following local spin frames for the four sublattices, labeled 0, 1, 2, and 3. All vectors are given in terms of Cartesian components in the crystallographic axes $\hat{a},\hat{b},\hat{c}$:
\begin{align}
\hat{x}_0 &= \frac{1}{\sqrt{6}}(-2,1,1);\quad
\hat{x}_1 = \frac{1}{\sqrt{6}}(-2,-1,-1);\quad
\hat{x}_2 = \frac{1}{\sqrt{6}}(2,1,-1);\quad
\hat{x}_3 = \frac{1}{\sqrt{6}}(2,-1,1);\nonumber\\
\hat{y}_0 &= \frac{1}{\sqrt{2}}(0,-1,1);\quad
\hat{y}_1 = \frac{1}{\sqrt{2}}(0,1,-1);\quad
\hat{y}_2 = \frac{1}{\sqrt{2}}(0,-1,-1);\quad
\hat{y}_3 = \frac{1}{\sqrt{2}}(0,1,1);\nonumber\\
\hat{z}_0 &= \frac{1}{\sqrt{3}}(1,1,1);\quad
\hat{z}_1 = \frac{1}{\sqrt{3}}(1,-1,-1);\quad
\hat{z}_2 = \frac{1}{\sqrt{3}}(-1,1,-1);\quad
\hat{z}_3 = \frac{1}{\sqrt{3}}(-1,-1,1);
\end{align}
In particular, $\hat{z}_i$ is parallel with the 3-fold axis that passes through site $i$.

The phase factors $\psi_{ij}$ arise due to the above local spin frames. $\psi_{ij} = \psi_{ji}$. Thanks to translation symmetry, $\psi_{ij}$ depends only on the sublattice labels of site $i$ and $j$. There are six independent $\psi_{ij}$ values:
\begin{align}
\psi_{01}=\psi_{23} = 0;\quad
\psi_{02}=\psi_{13} = \frac{2\pi}{3};\quad
\psi_{03}=\psi_{12}=-\frac{2\pi}{3}.
\end{align}

\section{Equations of motion of the toy model}

It is convenient to use the Lagrangian formalism. The Lagrangian of the toy model is given by:
\begin{subequations}
\begin{align}
L = \frac{\mathcal{N}}{4}\sum_{i}z_i \dot{\phi_i}-E,
\end{align}
where the energy
\begin{align}
E = -\frac{\mathcal{N}}{2}\sum_{i>j}\sqrt{(1-z^2_i)(1-z^2_j)}[\cos(\phi_i-\phi_j)-j_{\pm\pm}\cos(\phi_i+\phi_j+\psi_{ij})]+j_{zz}z_iz_j.
\end{align}
\end{subequations}
Here, $i$ and $j$ are sublattice indices running from 0 to 3. We have rescaled time such that $J_{\pm}S\to 1$. $j_{\pm\pm} \equiv J_{\pm\pm}/J_{\pm}$, and $j_{zz} \equiv J_{zz}/J_{\pm}$. $\mathcal{N}$ is the number of spins.

Using the new canonical variables $\phi$, $\phi_{a,b,c}$, $z$, and $z_{a,b,c}$ defined in the main text, and expanding $E$ to the quadratic order in all variables except for $\phi$, we find,
\begin{subequations}
\begin{align}
L = L_\phi+L_a+L_b+L_c,
\end{align}
where
\begin{align}
L_\phi &= \mathcal{N}z\dot{\phi}-3\mathcal{N}(1-j_{zz})z^2.\\
L_a & = \frac{\mathcal{N}}{4}z_a\dot{\phi}_a-\frac{\mathcal{N}}{4}(3+j_{zz})z^2_a-\mathcal{N}(1-\frac{j_{\pm\pm}}{2}\cos2\phi)\phi^2_a.\\
L_b & = \frac{\mathcal{N}}{4}z_b\dot{\phi}_b-\frac{\mathcal{N}}{4}(3+j_{zz})z^2_b-\mathcal{N}[1-\frac{j_{\pm\pm}}{2}\cos(2\phi+\frac{2\pi}{3})]\phi^2_b.\\
L_c & = \frac{\mathcal{N}}{4}z_c\dot{\phi}_c-\frac{\mathcal{N}}{4}(3+j_{zz})z^2_c-\mathcal{N}[1-\frac{j_{\pm\pm}}{2}\cos(2\phi-\frac{2\pi}{3})]\phi^2_c.
\end{align} 
\end{subequations}
Using Euler-Lagrangian equations, we obtain,
\begin{subequations}\label{eq:eom_all}
\begin{align}
\ddot{\phi}_a &=  -\omega^2_0(1-\frac{j_{\pm\pm}}{2}\cos2\phi)\phi_a.\\
\ddot{\phi}_b &=  -\omega^2_0[1-\frac{j_{\pm\pm}}{2}\cos(2\phi+\frac{2\pi}{3})]\phi_b.\\
\ddot{\phi}_c &=  -\omega^2_0[1-\frac{j_{\pm\pm}}{2}\cos(2\phi-\frac{2\pi}{3})]\phi_c.\\
\ddot{\phi} &=  -6j_{\pm\pm}(1-j_{zz})[\phi^2_a\sin2\phi+\phi^2_b\sin(2\phi+\frac{2\pi}{3})+\phi^2_c\sin(2\phi-\frac{2\pi}{3})].\label{eq:eom_phi}
\end{align}
\end{subequations}
Here, $\omega_0 \equiv 4\sqrt{3+j_{zz}}$ is a constant frequency. 

To derive an effective equation of motion for $\phi$, we approximate $\phi^2_{a,b,c}$ in Eq.~\eqref{eq:eom_phi} by their respective time average $\overline{\phi^2}_{a,b,c}$ in an oscillation cycle and, then, relate them to the action variables $I_{a,b,c}$:
\begin{align}
\overline{\phi^2_a} = \frac{I_a}{\omega_0\sqrt{1-j_{\pm\pm}/2\cos2\phi}};\quad
\overline{\phi^2_b} = \frac{I_b}{\omega_0\sqrt{1-j_{\pm\pm}/2\,\cos(2\phi+2\pi/3)}};\quad
\overline{\phi^2_c} = \frac{I_c}{\omega_0\sqrt{1-j_{\pm\pm}/2\,\cos(2\phi-2\pi/3)}}.
\end{align}
Substituting the above into Eq.~\eqref{eq:eom_phi}, we find an effective equation of motion for $\phi$:
\begin{subequations}
\begin{align}
\ddot{\phi} = -\frac{\partial V(\phi)}{\partial\phi},
\end{align}
where
\begin{align}
V(\phi) = \frac{3(1-j_{zz})}{\sqrt{3+j_{zz}}}\left( I_a\sqrt{1-\frac{j_{\pm\pm}}{2}\cos2\phi} + I_b\sqrt{1-\frac{j_{\pm\pm}}{2}\cos(2\phi+\frac{2\pi}{3})} + I_c\sqrt{1-\frac{j_{\pm\pm}}{2}\cos(2\phi-\frac{2\pi}{3})} \right).
\end{align}
\end{subequations}
This is the result given in the main text.

The above equation of motion for $\phi$ implies an \emph{effective} energy conservation law:
\begin{align}
\frac{E}{\mathcal{N}J_{\pm}S^2} = 3(1-j_{zz})z^2+\frac{V(\phi)}{6(1-j_{zz})} = \mathrm{constant}.
\end{align}
We recognize the first term as the kinetic energy of the pseudo-Goldstone mode and the second term as the effective potential energy due to the excited optical magnons.

\section{Modeling the magnetic pulse}

In this work, we model the magnetic pulse as a Dirac-$\delta$ function for the sake of simplicity: $\mathbf{B}(t) = B_\mathrm{max}\hat{n}\delta(t/\tau) = B_\mathrm{max}\tau\hat{n} \delta(t)$, where $\tau$ is the duration of the pulse, $B_\mathrm{max}$ is the peak field strength, and $\hat{n}$ is the unit polarization vector. In our simulation, we chose the pulse area $B_\mathrm{max}\tau = 0.1\mathrm{T}\cdot\mathrm{ps}$.

Using a Dirac-$\delta$ function is a sound approximation provided that the duration of the pulse $\tau$ is much shorter than oscillation period of the optical magnons. From Eq.~\eqref{eq:eom_all}, omitting a small dependence on $\phi$, the oscillation frequency of the optical magnon is $\omega_0 = 4\sqrt{3+j_{zz}}$ . Therefore, the oscillation period is approximately,
\begin{align}
\frac{2\pi\hbar}{4S \sqrt{J_{\pm}(3J_{\pm}+J_{zz})}} \approx \frac{2\pi\hbar}{0.5\mathrm{meV}} \approx 8\mathrm{ps}.
\end{align}
Here, we have restored the dimension of time. In the second equality, we have used the model parameters similar to Er$_2$Ti$_2$O$_7$. Setting a short duration $\tau = 1\mathrm{ps}$ and the peak strength typical for THz laser $B_\mathrm{max} = 0.1\mathrm{T}$ yields the peak area $B_\mathrm{max}\tau = 0.1\mathrm{T}\cdot\mathrm{ps}$.

We now analyze the impact of the Dirac $\delta$ pulse on spin dynamics. We begin with the toy model. The equations of motion are modified in the presence of the $\delta$ pulse:
\begin{subequations}
\begin{align}
\dot{\phi}_a &= 2n_a a_\parallel \delta(t);\quad \dot{z}_a = -2n_a a_\perp\sin\phi\cdot\delta(t);\\
\dot{\phi}_b &= 2n_b a_\parallel \delta(t);\quad \dot{z}_b = -2n_b a_\perp\sin(\phi-\frac{2\pi}{3})\cdot\delta(t);\\
\dot{\phi}_c &= 2n_c a_\parallel \delta(t);\quad \dot{z}_c = -2n_c a_\perp\sin(\phi+\frac{2\pi}{3})\cdot\delta(t);\\
\dot{\phi} &= 0;\quad \dot{z} = -\frac{a_\perp}{2}[n_a\phi_a\cos\phi+n_b\phi_b\cos(\phi-\frac{2\pi}{3})+n_c\phi_c\cos(\phi+\frac{2\pi}{3})]\delta(t).
\end{align}
\end{subequations}
Note that we have omitted the regular terms that do not involve $\delta(t)$. The dimensionless parameters $a_\perp \equiv 2g_\perp\mu_0 B_\mathrm{max}\tau/\sqrt{6}$ and $a_\parallel \equiv g_\parallel\mu_0 B_\mathrm{max}\tau/\sqrt{3}$. $n_{a,b,c}$ are the Cartesian components of the polarization vector $\hat{n}$ in the crystallographic (cubic) axis. Assuming $\phi_{a,b,c}(0^-) =\phi(0^-) = z_{a,b,c}(0^{-}) = z(0^-) = 0$ before the pulse, immediately after the pulse, we have
\begin{subequations}
\begin{align}
\phi_a(0^+) = 2n_a a_\parallel;\quad
\phi_b(0^+) = 2n_b a_\parallel;\quad
\phi_c(0^+) = 2n_c a_\parallel;\quad
\phi(0^+) = 0;\\
z_a(0^+) = 0;\quad
z_b(0^+) = \sqrt{3}n_b a_\perp;\quad
z_c(0^+) = -\sqrt{3}n_c a_\perp;\quad
z(0^+) = 0.
\end{align}
The second line may be rewritten as:
\begin{align}
\dot{\phi}_a(0^+) = 0;\quad
\dot{\phi}_b(0^+) = 2\sqrt{3} (3+j_{zz}) n_b a_\perp;\quad
\dot{\phi}_c(0^+) = -2\sqrt{3} (3+j_{zz}) n_c a_\perp;\quad
\dot{\phi}(0^+) = 0.
\end{align}
\end{subequations}
The above serve as the initial conditions for the ensuing free evolution govern by Eq.~\eqref{eq:eom_all}. In particular, one may evaluate the action variables $I_{a,b,c}$ by using the following:
\begin{align}
I_{a,b,c} = \frac{1}{2\omega_{a,b,c}}\dot{\phi}^2_{a,b,c} + \frac{\omega_{a,b,c}}{2}\phi^2_{a,b,c};\quad
\omega_a = \omega_0\sqrt{1-\frac{j_{\pm\pm}}{2}};\quad
\omega_b = \omega_0\sqrt{1+\frac{j_{\pm\pm}}{4}};\quad
\omega_c = \omega_0\sqrt{1+\frac{j_{\pm\pm}}{4}}.
\end{align}
Evidently, $I_{a,b,c}$ are respectively related to the values of $\phi_{a,b,c}$ and $\dot{\phi}_{a,b,c}$, which in turn depend on the polarization vector $\hat{n}$.

In simulating the full lattice model, the Dirac-$\delta$ magnetic pulse is treated in the same spirit. Let $\mathbf{S}_i = S^x_i\hat{x}_i+S^y_i\hat{y}_i+S^z_i\hat{z}_i$ be the spin vector on a pyrochlore lattice site $i$. $S^{x,y,z}_i$ are the Cartesian components in the local spin frame $\{\hat{x}_i,\hat{y}_i,\hat{z}_i\}$. The pulse induces a pure rotation:
\begin{align}
\left(\begin{array}{c}
S^x_i(0^+)\\
S^y_i(0^+)\\
S^z_i(0^+)
\end{array}\right)
= R(\frac{\mathbf{h}}{h},h\tau)
\left(\begin{array}{c}
S^x_i(0^-)\\
S^y_i(0^-)\\
S^z_i(0^-)
\end{array}\right).
\end{align}
Here, $S^{x,y,z}_i(0^\pm)$ are the Cartesian components immediately after and before the pulse. $R(\hat{u},\theta)$ is the $3\times 3$ Euclidean rotation matrix with respect to the rotational axis $\hat{u}$ by an angle $\theta$. The vector $\mathbf{h} = (g_\perp\mu_0 B \hat{n}\cdot\hat{x}_i, g_\perp\mu_0 B \hat{n}\cdot\hat{y}_i, g_\parallel\mu_0 B \hat{n}\cdot\hat{z}_i)$ is the Zeeman field in the local spin frame, and $h$ is its length. 

\section{Monte Carlo simulation}\label{MC}

\begin{figure}[t]
\includegraphics[width=\columnwidth]{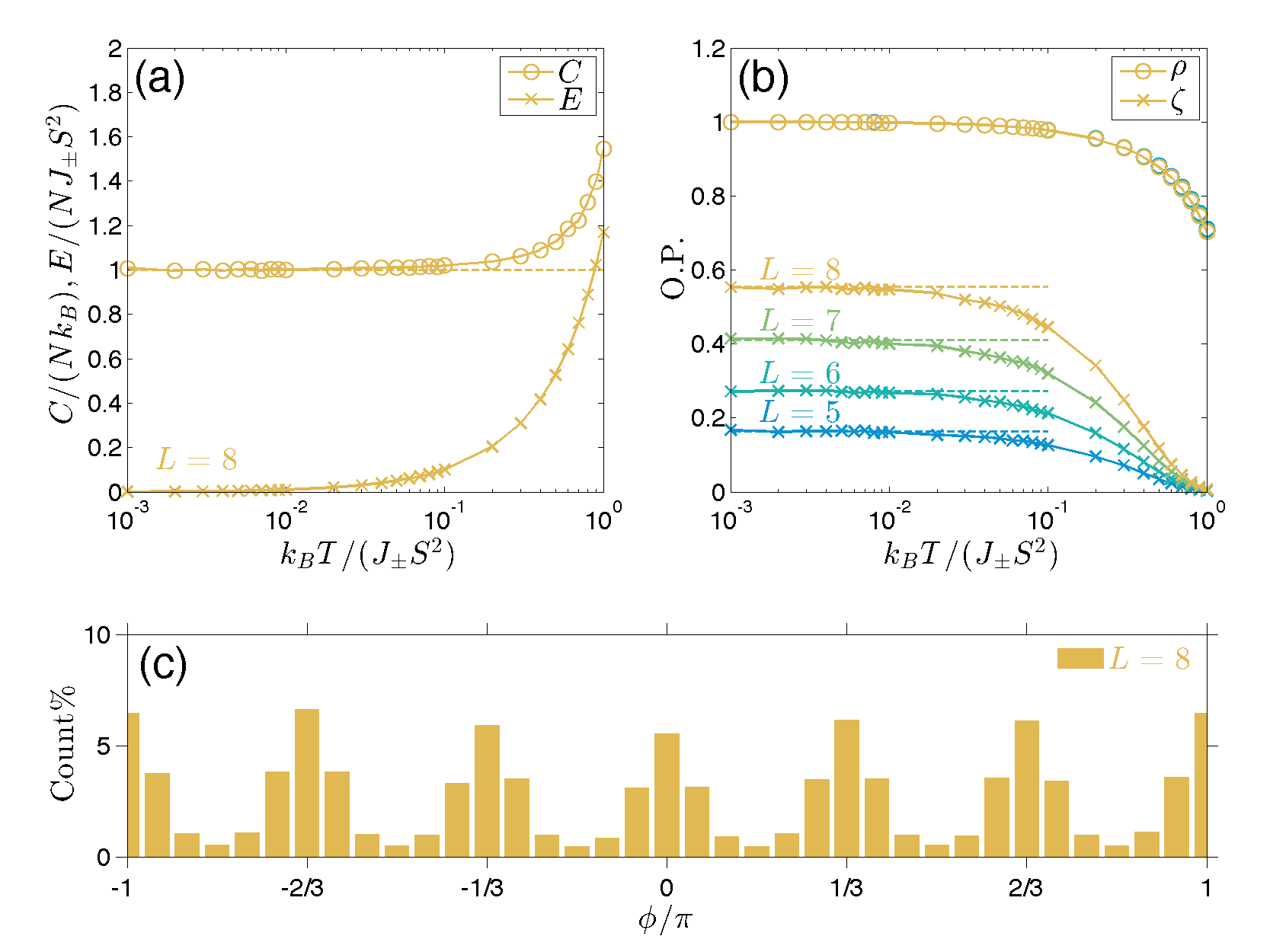}
\caption{(a) Energy (crosses) and specific heat (circles) per spin as function of temperature $T$ from MC simulation on a system of $L\times L\times L$ unit cells. Here, results for $L=8$ are shown. Dashed line marks the predicted specific heat ($k_B$ per spin) from equipartition in the low temperature limit. (b) Order parameters $\rho$ (circles) and $\zeta$ (crosses) versus $T$, for systems of various linear dimension $L$. $\rho$ characterizes the N\'{e}el order, whereas $\zeta$ characterizes the six-state clock order (see Sec.~\ref{MC} for their definition). Dashed lines mark the value of $\lim_{T\to0}\zeta$ predicted by linear spin wave theory. Data for different $L$s are shown in different colors. (c) Histogram of the N\'{e}el order parameter argument $\phi$ at $k_\mathrm{B}T=10^{-2}J_{\pm}S^2$ in a system of linear dimension $L=8$. $\phi$-axis ticks mark the $\phi$ values that correspond to the $\psi_2$ states. 36 equal-size bins are used to obtain the histogram.}
\label{fig:thermal}
\end{figure}

In our finite temperature molecular dynamics simulation, we use canonical Monte Carlo (MC) method to generate initial conditions for the Landau-Lifshitz equation. In the MC simulation, we equilibrate the system with $2^{15}$ MC steps. We skip $2^6$ MC steps before we record a new configuration as a sample in order to reduce the sample correlation. In thermal equilibrium, the six $\psi_2$ states occur with equal probability, giving rise to six magnetic domains. We assume the system is initially in a single domain of the $\psi_2$ states with $\phi=0$. This is done by discarding samples whose order parameter argument $\mathrm{arg}\,O$ falls out of the interval $\mathrm{arg}\,O\in[-\pi/6,\pi/6)$. We generate $2^{10}$ samples and feed them to the Runge-Kutta integrator as the initial conditions. 

The details of the MC update scheme are given as follows. Each MC step consists of 1 spin flip update and 8 over-relaxation updates:

\begin{itemize}
\item To flip a spin, we draw with uniform probability a new orientation from a spherical cap that centers on the previous orientation of the spin. This procedure is performed on each and every spin in a spin flip update. The size of the cap is a parameter of the update, which we adjust to reduce the autocorrelation time. The updated configuration is accepted according to the Metropolis rule. 

\item An over-relaxation update consists of four steps of rotations. In each step, we rotate each and every spin belonging to a given sublattice $i$ with respect to its local molecular field by angle $\pi$. The rotations within the \emph{same} sublattice are independent from each other. As a result, the precise order of the rotations within a sublattice does not affect the outcome. However, rotations for two \emph{different} sublattices do not commute. We first rotate the spins belonging to sublattice 0, and then proceed to sublattice 1, 2, and 3. The updated configuration is always accepted. By construction, the over-relaxation update conserves energy and satisfies detailed balance. 
\end{itemize}

We benchmark our MC update scheme by measuring the low temperature thermodynamic properties of the model. To do this, we perform a separate MC simulation. The results are summarized in Fig.~\ref{fig:thermal}, where the data are obtained by averaging over $2^{20}$ successive MC steps after equilibrating the system for $2^{18}$ MC steps. Note we no longer discard MC samples that fall outside of the $\psi_2$ domain with $\phi=0$. 

In agreement with the equipartition theorem, the low temperature specific heat clearly approaches $k_B$ (Fig.~\ref{fig:thermal}a). To characterize thermal order by disorder, we use two order parameters: $\rho \equiv \sqrt{\langle |O|^2\rangle}$, and $\zeta \equiv \langle\cos6\phi\rangle$, where $|O|$ and $\phi$ are respectively the modulus and argument of the complex N\'{e}el order parameter $O$, and $\langle\cdots\rangle$ denotes the ensemble average. $\rho$ characterizes the N\'{e}el order, whereas $\zeta$ characterizes the six-state clock order. In particular, $\zeta = 1$ if the system falls precisely on a $\psi_2$ state. Fig.~\ref{fig:thermal}b shows the N\'{e}el order is well established in the temperature window $k_BT<J_{\pm}S^2$. As $T$ decreases, the six-state clock order starts to develop and quickly approaches the value predicted by linear spin wave theory. 

While $\rho$ exhibits little dependence on system size (hence the $\rho$ plots of different $L$ lie on top of each other), $\zeta$ shows significant finite size effect. The latter is due to the entropic nature of the six-state clock anisotropy. In the $T\to0$ limit, the probability distribution of the order parameter argument $\phi$ is given by $\rho(\phi)\propto \exp[\mathcal{N}s(\phi)]$, where $s(\phi)$ is the thermal entropy density due to spin wave fluctuations and $\mathcal{N}$ is the number of spins. The peaks of $\rho(\phi)$ are located at $\psi_2$ states (Fig.~\ref{fig:thermal}c). Crucially, for finite but large $\mathcal{N}$, the width of these peaks is inversely proportional to $\sqrt{\mathcal{N}}$. As $\mathcal{N}$ increases, more and more spin configurations fall on $\psi_2$ states, and hence $\zeta$ increases towards 1.

\begin{figure}
\includegraphics[width=\textwidth]{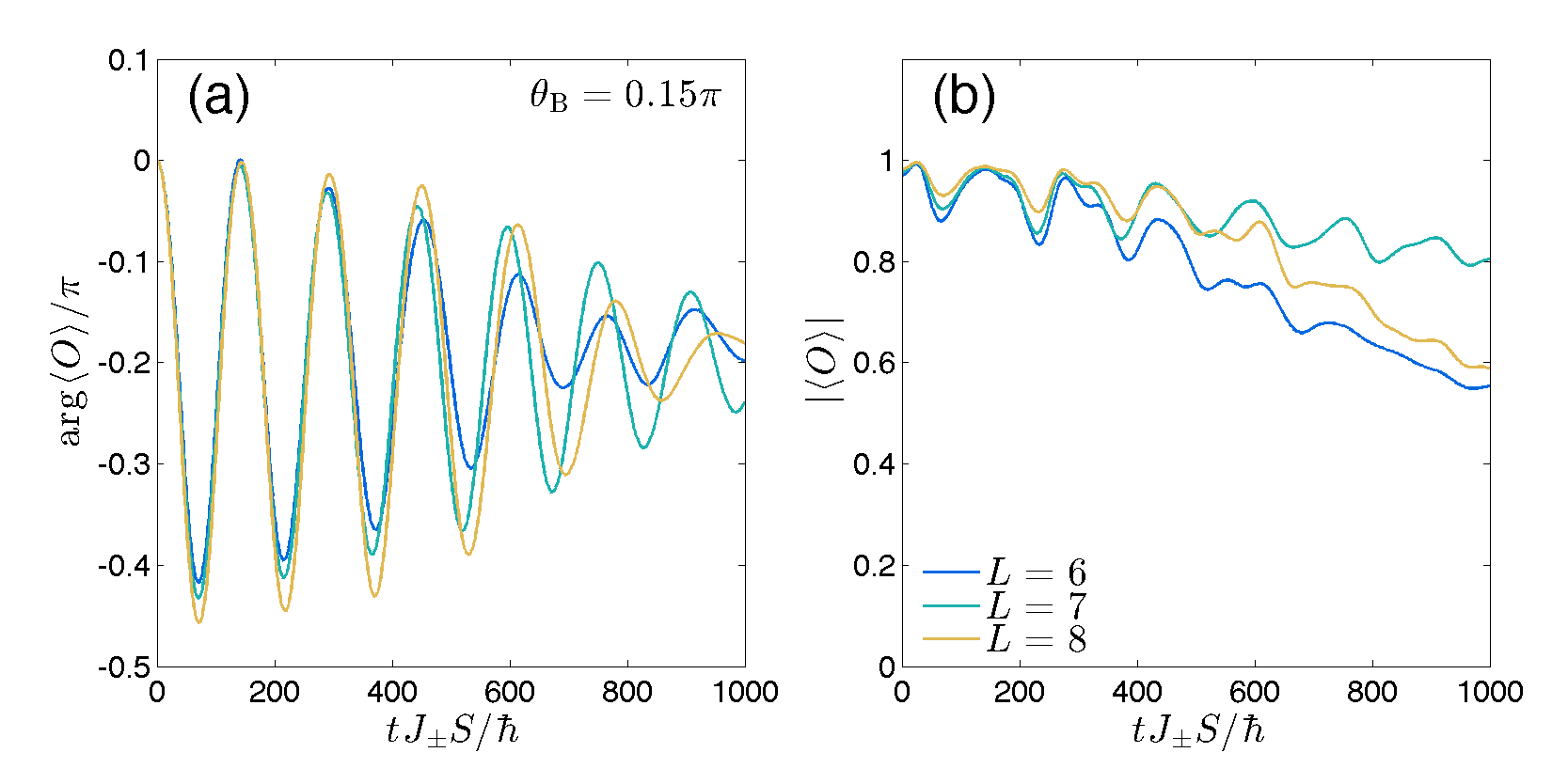}
\caption{Argument (a) and modulus (b) of the ensemble-averaged N\'{e}el order parameter $\langle O\rangle$ as a function of time $t$ after a pulse. The pulse polarization angle $\theta_\mathrm{B}=0.15\pi$,  and temperature $k_BT = 10^{-3}J_{\pm}S^2$. Results for three lattice sizes are shown in different colors.} 
\label{fig:finitesize}
\end{figure}

\section{Finite size effect of the order parameter dynamics}

In our simulation, we use a system of $L\times{}L\times{}L$ unit cells with linear dimension $L=8$. Here, we assess the finite size effect of the order parameter dynamics by monitoring its dependence on $L$. Fig.~\ref{fig:finitesize} shows the argument and the modulus of the average N\'{e}el order parameter $\langle O\rangle$ as a function of time for $L=6,7,8$. The temperature $k_BT=10^{-3}J_{\pm}S^2$. The pulse polarization angle $\theta_\mathrm{B}=0.15\pi$. 

For all three linear dimensions, $\mathrm{arg}\langle{}O\rangle$, the argument of the average N\'{e}el order parameter $\langle{}O\rangle$, exhibit qualitatively similar behavior within the time window of simulation. At early times, the oscillation amplitude of $\mathrm{arg}\langle{}O\rangle$ increases rather than decreases with $L$. Meanwhile, the center position of the oscillation, which indicates the position of a non-equilibrium potential minimum, changes with $L$ by a small amount. Likewise, the modulus of the average order parameter, $|\langle O\rangle|$, decays in time for all three $L$s. We also note that the decay rate of the modulus $|\langle O\rangle|$ shows non-monotonic dependence on $L$: the slope of the $L=7$ plot is greater than those of $L=6$ and $L=8$.

Without a finite-size scaling theory, we are unable to perform controlled extrapolation toward $L\to\infty$. Nevertheless, given that the order parameter dynamics shows only small dependence on $L$ within the simulated time window, the qualitative features of the non-equilibrium dynamics should be robust in the thermodynamic limit.

\end{document}